\documentclass[english,pre,11pt,twolumn]{revtex4}
\usepackage{amsfonts}
\usepackage{amsmath}
\usepackage{graphicx}
\usepackage{color}
\usepackage{babel}
\usepackage{hyperref}
\usepackage{epsfig,bm}
\usepackage{graphicx}
\usepackage{soul}
\usepackage[normalem]{ulem}
\usepackage{enumerate}

\begin{document}
\title{A scaling law beyond Zipf's law and its relation to Heaps' law}
\author{Francesc Font-Clos}
\affiliation{Centre de Recerca Matem\`atica, Edifici C, Campus Bellaterra, E-08193 Bellaterra (Barcelona), Spain}
\affiliation{Departament de Matem\`atiques, Universitat Aut\`onoma de Barcelona, Edifici C, E-08193 Bellaterra (Barcelona), Spain}
\author{Gemma Boleda}
\affiliation{Department of Linguistics, 
The University of Texas at Austin, 
1 University Station B5100,
Austin, TX, USA}
\author{\'Alvaro Corral}
\affiliation{Centre de Recerca Matem\`atica, Edifici C, Campus Bellaterra, E-08193 Bellaterra (Barcelona), Spain}

\begin{abstract}
The dependence on text length of the statistical properties of 
word occurrences
 has long been considered a severe limitation
{on the usefulness of}
quantitative linguistics.
We propose a simple scaling form for the distribution
of absolute word frequencies
that brings to light the robustness of this distribution
as text grows. In this way, the shape of the distribution
is always the same, and it is only a scale parameter that 
increases (linearly) with text length.
By analyzing very long novels we show that
this behavior holds both for raw, unlemmatized texts and for lemmatized texts.
In the latter case, the distribution of frequencies is well
approximated by a double power law, maintaining the Zipf's exponent value
$\gamma\simeq 2$
for large frequencies but yielding a smaller exponent
in the low-frequency regime.
The growth of the distribution with text length allows us to
estimate the size of the vocabulary at each step and to propose
a generic alternative to Heaps' law,
which turns out to be intimately connected to 
the distribution of frequencies,
thanks to its scaling behavior.

\end{abstract}

\maketitle
\section{Introduction}

Zipf's law is perhaps one of the best pieces of evidence
about the existence of universal physical-like laws
in cognitive science and the social sciences.
Classic examples where it applies include the population of cities, 
company income,
and the frequency of words in texts or speech \cite{Zipf_1949}.
In the latter case,
the law is obtained directly by counting the number of repetitions, 
i.e., the absolute frequency $n$, of all words in a long enough text, 
and assigning increasing ranks, $r=1,2, \dots$, to decreasing frequencies. 
When a power-law relation
$$
n \propto \frac 1 {r^\beta}
$$
holds for a large enough range,
with the exponent $\beta$ more or less close to 1,
Zipf's law is considered to be fullfilled
(with $\propto$ denoting proportionality).
An equivalent formulation of the law is obtained 
in terms of the probability distribution of the frequency $n$, 
such that it plays the role of a random variable, for which a power-law distribution
$$
D(n) \propto \frac 1 {n^\gamma},
$$
should hold, with $\gamma=1+1/\beta$ (taking values close to 2) and
$D(n)$ as the probability mass function of $n$ (or the probability density of $n$, 
in a continuous approximation) \cite{Mandelbrot61,Ferrer2008f,Adamic_Huberman,Kornai2002,Zanette_book}. 
Note that this formulation implies performing
double statistics (i.e., doing statistics twice),
first counting words to get frequencies and then counting repetition of frequencies to get the distribution
of frequencies.

The criteria for the validity of Zipf's law are arguably rather vague 
(long enough text, large enough range, exponent $\beta$ more or less close to 1). 
Generally, a long enough text means a book, 
a large range can be a bit more than an order of magnitude, 
and the proximity of the exponent $\beta$ to 1 translates into an interval (0.7,1.2),
or even beyond that
\cite{Zanette_2005,Zanette_book,Ferrer2004a}.
Moreover, no rigorous methods have been usually required for the fitting of the power-law distribution:
Linear regression in double-logarithmic scale is the most common method,
either for $n(r)$ or for $D(n)$,
despite the fact that it is well known that this procedure suffers from severe drawbacks and can lead to flawed results
\cite{Clauset,Corral_nuclear}.
Nevertheless, once these limitations are assumed, the fulfillment of Zipf's law
in linguistics is astonishing, being valid no matter the author, style, or language
\cite{Zipf_1949,Zanette_2005,Zanette_book}.
So, the law is universal, at least in a qualitative sense.

At a theoretical level, many different competing explanations of Zipf's law have been proposed \cite{Zanette_book}, 
such as random (monkey) typing \cite{Miller_monkey,Li_1992}
,preferential repetitions or proportional growth \cite{Simon,Newman_05,Saichev_Sornette_Zipf},
the principle of least effort \cite{Zipf_1949,Ferrer2002a,Corominas-Murtra_2011,Ferrer2004e},
and, beyond linguistics,
Boltzmann-type approaches \cite{Toscani_2008},
or even avalanche dynamics in a critical system \cite{Bak_book};
most of these options have generated considerable controversy
\cite{Mitz,Ferrer-i-Cancho_2010,Dickman_Moloney_Altmann}.
In any case, the power-law behavior is the hallmark of scale invariance,
i.e., the impossibility to define a characteristic scale, 
either for frequencies or for ranks.
Although power laws are sometimes also referred to as scaling laws, 
we will make a more precise distinction here.
In short, a scaling law is any function invariant under a scale transformation
(which is a linear dilation or contraction of the axes).
In one dimension the only scaling law is the power law, 
but this is not true with more than one variable \cite{Christensen_Moloney}. Note that 
in text statistics, other variables to consider in addition to frequency are the
text length $L$ (the total number of words, or tokens) and the size of the vocabulary 
$V_L$ (i.e., the number of different words, or types).

%

Somehow related to Zipf's law is Heaps' law (also called Herdan's law \cite{Baayen,Herdan_book}), which states that the vocabulary $V_L$ 
grows as a function 
of the text length $L$ as a power law,
$$
V_L \propto L^{\alpha},
$$ 
with the exponent $\alpha$ smaller than one. However, even simple log-log plots of $V_L$ versus $L$ do not show a
convincing linear behavior
\cite{Altmann2012}
 and 
 therefore,
the evidence for this law is somewhat weak
(for a notable exception see Ref. \cite{Kornai2002}). 
Nevertheless, 
a number of works have derived the relationship $\beta=1/\alpha$
between Zipf's and Heaps' exponents \cite{Mandelbrot61,Kornai2002,Leijenhorst_2005},
at least in the infinite-system limit \cite{Serrano_2009, Lu_2010},
using different assumptions.

Despite the relevance of Zipf's law, and its possible relations with criticality,
few systematic studies about the 
dependence of the law on system size (i.e., text length) have been carried out.
It was Zipf himself \cite[pp. 144]{Zipf_1949} who first observed 
a variation in the exponent $\beta$ when the system size was varied. 
In particular, ``small'' samples would give $\beta<1$, while ``big'' ones yielded $\beta > 1$.
However, that was attributed to ``undersampling'' and 
``oversampling'', as Zipf believed that there was an optimum system size 
under which all words occurred in proportion to their theoretical frequencies, i.e., 
those given by the exponent $\beta=1$.
This increase of $\beta$ with $L$ has been confirmed later, see Refs.~\cite{Powers_1998, Baayen}, 
leading to the conclusion that the practical usefulness of Zipf's law is rather limited \cite{Baayen}.

More recently, using rather large collections of books from 
single authors,
Bernhardsson et al. \cite{Bernhardsson_2009}
find
a decrease of the exponents $\gamma$ and $\alpha$
with text length, in correspondence with the increase in $\beta$ found by Zipf and others. 
They propose a size-dependent word-frequency distribution based on three main assumptions:

\begin{enumerate}[(i)]
\item The vocabulary scales with text length as $V_L \propto L^{\alpha(L)}$, 
where the exponent $\alpha (L)$ itself depends on the text length.
Note however that this is not an assumption in itself, just notation,
and it is also equivalent 
to writing
the average frequency 
$\langle n \rangle = L/V_L $ 
as $\langle n (L)\rangle \propto L^{1-\alpha(L)}$.

\item The maximum frequency is proportional to the text length, 
i.e. $n_{\max} = n (r=1) \propto L$. 
\item The functional form of the word frequency distribution $D_L(n)$ is that 
of a power law with an exponential tail, 
with both the scale parameter $c(L)$ and the power-law exponent $\gamma(L)$  
depending on the text length $L$. That is,
\[
D_L (n) =A \frac{e^{- n/c(L)}}{n^{\gamma(L)}},
\]
with $1 < \gamma(L) < 2$.
\end{enumerate}

Taking $c(L)=c_0 L$ guarantees that $n_{\max} \propto L$;
moreover, the form of $D_L(n)$ implies that, asymptotically, 
$\langle n (L)\rangle \propto L^{2-\gamma(L)}$ \cite{Christensen_Moloney},
which comparing to assumption (i)
leads to 
$$
\alpha(L)=\gamma(L)-1,
$$
so, $0<\alpha(L)<1$. This relationship between $\alpha$ and $\gamma$ is in agreement with previous results {if $L$ is fixed} 
\cite{Mandelbrot61,Lu_2010,Serrano_2009}. 
It was claimed in Ref. \cite{Bernhardsson_2009}
that $\alpha(L)$ decreases from 1 to 0
for increasing $L$ and therefore $\gamma(L)$ decreases from 2 to 1. The resulting functional form,
\[
D_L (n) =A \frac{e^{- n/(c_0 L)}}{n^{1+\alpha(L)}},
\]
 is in fact the same functional form appearing in many critical phenomena,
where the power-law term is limited by a characteristic value of the variable, $c_0 L$,
arising from a deviation from criticality or from finite-size effects
\cite{Christensen_Moloney,Aharony,Zapperi_branching,Corral_FontClos}. 
Note that 
this implies that the tail of the frequency distribution 
is not a power law but an exponential one,
and therefore the frequency of most common words
is not power-law distributed.
This is in contrast with recent studies that have 
clearly established that the tail of $D_L(n)$
is well modelled by a power law \cite{Clauset,Corral_Boleda}.
%
However, what is most uncommon about this functional form is the fact that it has a ``critical'' 
exponent that depends on system size:
The values of exponents should not be influenced by external scales.
So, here we look for an alternative picture that is more in agreement 
with typical scaling phenomena.

Our proposal is that, although the word-frequency distribution
$D_L(n)$ changes with system size $L$, 
the \emph{shape} of the distribution is independent
of $L$ and $V_L$, 
and only the \emph{scale} of $D_L(n)$ changes with 
these variables.
This implies that the shape parameters of $D_L(n)$ (in particular,
any exponent) do not change with $L$; only one scale parameter changes with $L$, 
increasing linearly.
This is explained in the next section, while the third one
is devoted to the validation of our scaling form in real texts, 
 using both plain words and their corresponding lemma forms;
in the latter case an alternative to Zipf's law can be proposed, 
consisting of a double power-law distribution
(which is a distribution with two power-law regimes that have
different exponents).
Our findings for words and lemmas suggest that the previous observation that the exponent in Zipf's law depends on text length 
\cite{Powers_1998, Baayen, Bernhardsson_2009},
might be an artifact of the increasing weight of a second regime in the distribution of frequencies beyond a certain text length.
The fourth section investigates the implications of our scaling approach
for Heaps' law.
Although the scaling ansatz we propose has a counterpart in the rank-frequency representation,
we prefer to illustrate it in terms of the distribution of frequencies, as this approach
has been deemed more appropriate from a statistical point of view \cite{Corral_Boleda}.


\section{The scaling form of the word-frequency distribution}

Let us come back to the rank-frequency relation, in which 
the absolute frequency $n$ of each type is a function of its rank $r$.
Defining the relative frequency as $x\equiv n/L$
and inverting the relationship, we can write
$$
r=G_L(x).
$$
Note that here we are not assuming a power-law relationship
between $r$ and $x$, just a generic function $G_L$, which may depend
on the text length $L$.
Instead of the three assumptions introduced by Bernhardsson et al.
we just need one assumption, which is the independence of the function
$G_L$ with respect to $L$;
so 
\begin{equation}
r=G(n/L).
\label{rGnL}
\end{equation}
This turns out to be a scaling law, with $G(x)$ a scaling function.
It means that if in the first 10,000 tokens of a book there are 
5 types with relative frequency 
larger than or equal to 2\%, that is, $G(0.02)=5$,
then this will still be true for the first 20,000 tokens, 
and for the first 100,000 and for the whole book. 
These types need not necessarily be the same ones, although in some cases they might be.
In fact, instead of assuming as in Ref. \cite{Bernhardsson_2009}
that the frequency of the most used type scales linearly
with $L$, what we assume is just that this is true for all types, at least on average.
Notice that this is not a straightforward assumption, 
as, for instance, Ref. \cite{Kornai2002} considers instead that $n$ is just a (particular) function of $r/V_L$.

Now let us introduce the survivor function or complementary cumulative distribution function
$S_L(n)$ of the absolute frequency, defined in a text of length $L$ as 
$S_L(n)=\mbox{Prob}[\mbox{frequency} \ge n]$.
Note that, estimating from empirical data, 
$S_L(n)$ turns out to be essentially the rank, but divided by the total number of ranks, 
$V_L$, i.e., $S_L(n)=r/V_L$. Therefore, using our ansatz for $r$ we get
$$
S_L(n)=\frac{G(n/L)}{V_L}.
$$
Within a continuous approximation the probability mass function of $n$,
 $D_L(n)=\mbox{Prob}[\mbox{frequency} = n]$, can be
obtained from the derivative of $S_L(n)$,
\begin{equation}
\label{main}
D_L(n) = -\frac{\partial S_L(n)}{\partial n}= \frac{g(n/L)}{L V_L},
\end{equation}
where $g$ is minus the derivative of $G$, i.e., $g(x)=-G\hspace*{0.7mm}'(x)$.
If one does not trust the continuous approximation, 
one can write $D_L(n)=S_L(n)-S_L(n+1)$ and perform a 
Taylor expansion, for which the result is the same, 
but with $g(x)\simeq -G\hspace*{0.7mm}'(x)$.
In this way, we obtain simple forms for $S_L(n)$ and $D_L(n)$,
which are analogous to standard scaling laws, except for the fact that
we have not specified how $V_L$ changes with $L$.
If Heaps' law holds, $V_L\propto L^\alpha$, we recover a standard scaling law,
$D_L(n)=g(n/L)/L^{1+\alpha}$,  
which fulfills invariance under a scaling
transformation, 
or, equivalently, fulfills the definition
of a generalized homogeneous function \cite{Christensen_Moloney, Stanely_1972}, 
$$
D_{\lambda_L L}(\lambda_n n) = \lambda_D D_L(n),
$$ 
where $\lambda_L$, $\lambda_n$, and $\lambda_D$ are the scale
factors, related in this case through
$$
\lambda_n=\lambda_L \equiv \lambda
$$
and
$$
\lambda_D=\frac 1 {\lambda^{1+\alpha}}.
$$
However, in general (if Heaps' law does not hold),
the distribution $D_L(n)$ still is invariant under a scale
transformation but with a different relation for $\lambda_D$, which is
$$
\lambda_D=\frac {V_L}{\lambda V_{\lambda L}}.
$$
So, $D_L(n)$
is not a generalized homogeneous function,
but presents an even more general form.
In any case,
the validity of the proposed scaling law, Eq. (1), can be checked by performing a very simple 
rescaled plot, displaying $L V_L D_L(n)$ versus $n/L$.
A resulting data collapse support the independence of the scaling
function with respect to $L$.  This is undertaken in the next section.

%


%

\section{Data analysis results}


To test the validity of our predictions, 
summarized in Eq.~\eqref{main}, 
we analyze a corpus of literary texts,
comprised by seven large books in English, Spanish, and French
(among them, some of the longest novels ever written,
in order to have as much statistics of homogeneous texts as possible).
In addition to the statistics of the words in the texts,
we consider the statistics of lemmas
(roughly speaking, the stem forms of the word; for instance, \textit{dog} for \textit{dogs}).
In the lemmatized version of each text,
each word is substituted by its corresponding lemma,
and the statistics are collected in the same way as they are collected for word forms.
Appendix~\ref{appendix lemmatization} provides detailed information on the lemmatization procedure, and
Table \ref{lengths} summarizes the
most relevant characteristics of the analyzed books.

\begin{table}[h]
\centering
\begin{tabular}{|l|l|l|r|r|r|r|r|}
\hline
 Title			& author 	& language 	& year	&$L_{\text{tot}}$ 	& $V_{\text{tot}}$	& $L_{\text{tot}}^{(l)}$ 	& $V_{\text{tot}}^{(l)}$ \\
\hline
Artam\`ene   	& Scud\'ery siblings 	& French & 1649 	&2,078,437	&25,161 		&1,737,556	&5,008	\\
\hline
 Clarissa        	& Samuel Richardson 	& English		& 1748	&971,294		&20,490 		&940,967		&9,041	\\
\hline
 Don Quijote	& Miguel de Cervantes	& Spanish	& 1605-1615 	&390,436		&21,180 		&378,664		&7,432	\\
\hline
 La Regenta	& L. Alas ``Clar\'in''	& Spanish	& 1884	&316,358		&21,870		&309,861		&9,900	\\
\hline
 Le Vicomte de Bragelonne  	& A. Dumas (father) & French & 1847 &693,947		&25,775		&676,252		&10,744	\\
\hline
 Moby-Dick	& Herman Melville	& English	& 1851	&215,522		&18,516 		&204,094		&9,141	\\
\hline
 Ulysses        	& James Joyce		& English	& 1918	&268,144		&29,448		&242,367		&12,469	\\
\hline
\end{tabular}
\caption{
Total text length and vocabulary before ($L_{\text{tot}}, V_{\text{tot}}$) and after ($L_{\text{tot}}^{(l)}, V_{\text{tot}}^{(l)}$) 
the lemmatization process, for all the books considered (including also their author, language, and publication year). 
The text length for lemmas is shorter than for words because for a number of word tokens
their corresponding lemma type could not be determined, and they were ignored.
}
\label{lengths}
\end{table}

\begin{figure}
\centering
\includegraphics[width=\textwidth]{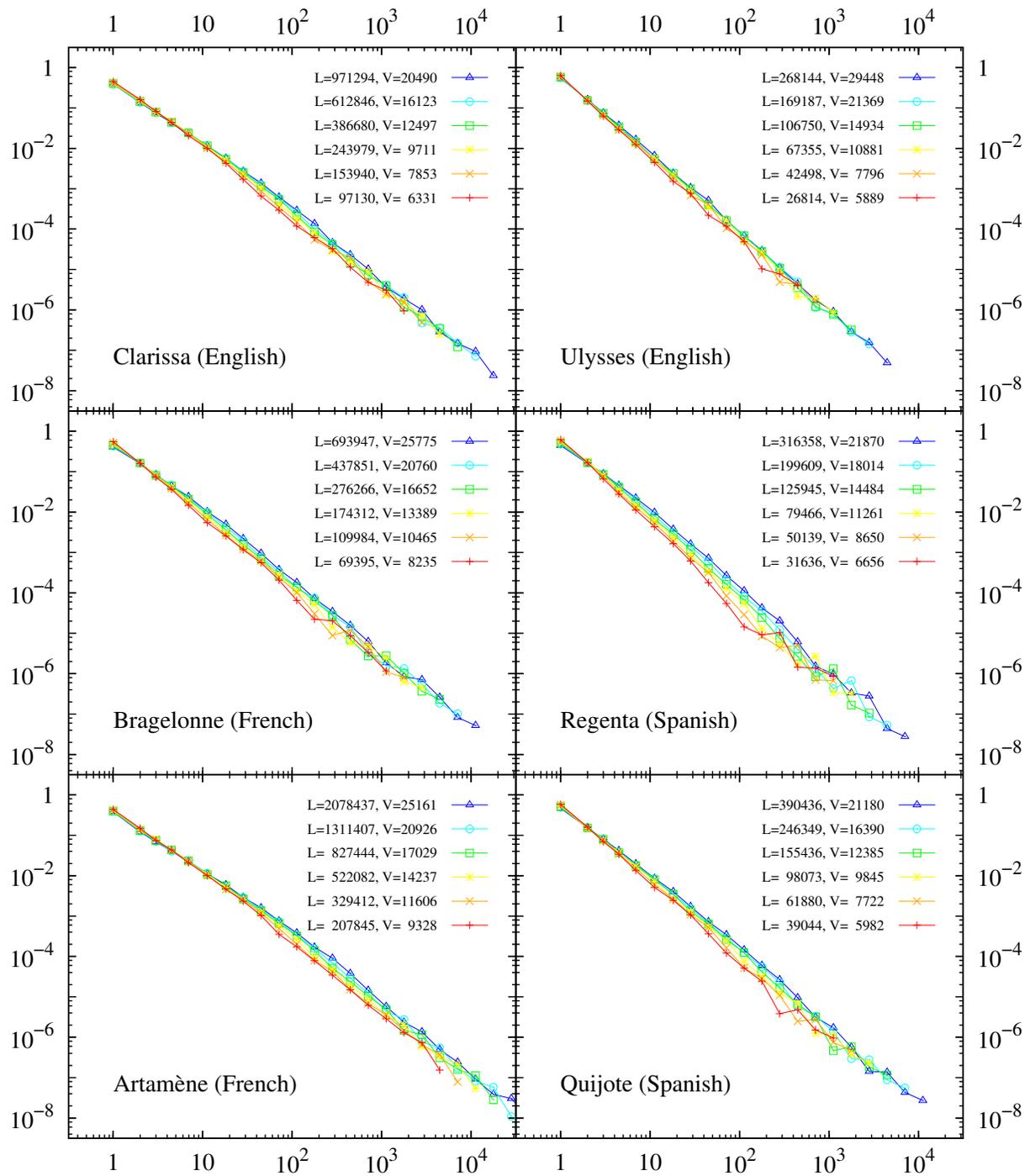}
\caption{
Density of word frequencies $D_L(n)$ ($y$-axis) against absolute frequency $n$ ($x$-axis), 
for six different books, taking text length $L=L_{\text{tot}}/10$, $L_{\text{tot}}/10^{4/5}$, $L_{\text{tot}}/10^{3/5}$, $\dots$,
$L_{\text{tot}}$. The slope seems to decrease with text length.
}
\label{data no collapse}
\end{figure}

First,
we plot the distributions of word frequencies, $D_L(n)$ versus $n$, for each book, 
considering either the whole book or the first $L/L_{\text{tot}}$ fraction,
where $L_{\text{tot}}$ is the real, complete text length
(i.e., if $L=L_{\text{tot}}/2$ we consider just the first half of the book, 
no average is performed over parts of size $L$). 
For a fixed book, we observe that different $L$ leads to 
distributions with 
small but clear
differences, see Fig.~\ref{data no collapse}.
The pattern described by Bernhardsson et al. (equivalent to Zipf's findings for the change of the exponent $\beta$) seems to hold, as the 
absolute value of the slope in log-log scale (i.e., the apparent power-law exponent $\gamma$)
decreases with increasing text length.

\begin{figure}
\centering
\includegraphics[width=\textwidth]{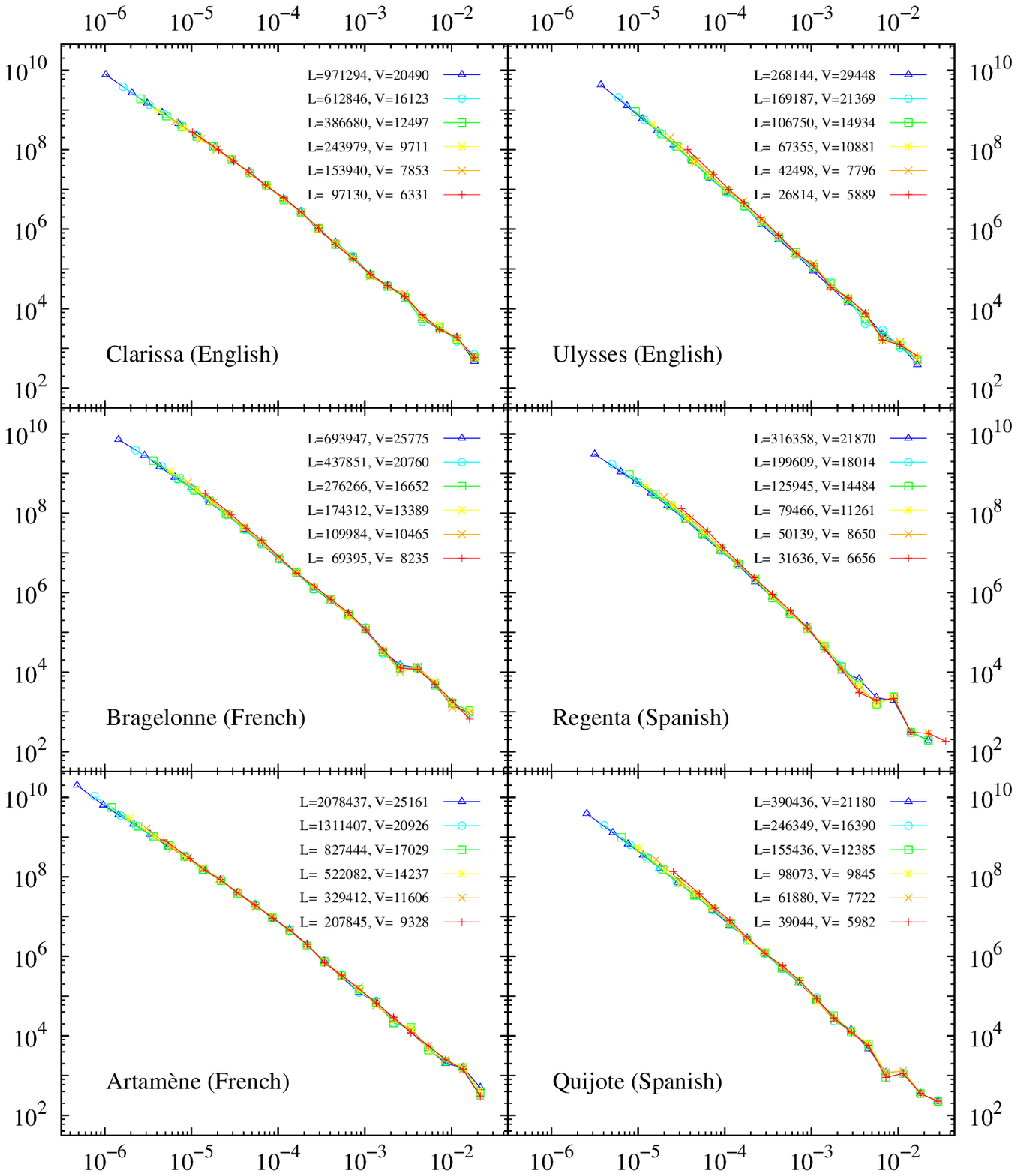}
\caption{
Rescaled densities $L V_L D_L(n) $ ($y$-axis) against relative frequency $n/L$ ($x$-axis), 
for the same books and fractions of text as in Fig.~\ref{data no collapse}. 
The rescaled densities collapse onto a single function, independently of the value of $L$, 
validating our proposed scaling form for $D_L(n)$
 [Eq. \eqref{main}] and making it clear that the decrease of the log-log slope with $L$ is 
not a consequence of a genuine change in the scaling properties of the distribution.}
\label{data collapse}
\end{figure}

However, a scaling analysis reveals an alternative picture.
As suggested by Eq.~\eqref{main}, plotting $LV_L D_L(n)$ against $n/L$ for different values of $L$ 
yields a collapse of all the curves onto a unique $L-$independent function for each book, 
which represents the scaling function $g(x)$.
Figure \ref{data collapse} shows this for 
the same books and parts of the books as in Fig.~\ref{data no collapse}.
The data collapse can be considered excellent, except for the smallest frequencies.
 For the largest $L$ the collapse is valid up to $n\simeq3$
if we exclude {\it La Regenta}, which only collapses for about $n \ge 6$.
So, our scaling hypothesis is validated,
independently of the particular shape that $g(x)$ takes. 
Note that $g(x)$ is independent of $L$ but not the book, 
i.e., each book has its own $g(x)$, 
different from the rest.
%
In any case, we observe a slightly convex shape in log-log space, 
 which leads to the rejection of the power-law hypothesis for the whole range of frequencies.
Nevertheless,
the data does not show any clear parametric functional form. 
A double power law, a stretched exponential, a Weibull, or a lognormal tail
could be fit to the distributions.
This is not incompatible with the fact that the large $n$ tail
can be well fit by a power law (the Zipf's law),
for more than 2 orders of magnitude \cite{Corral_Boleda}.

\begin{figure}[ht!]
\includegraphics[width=\textwidth]{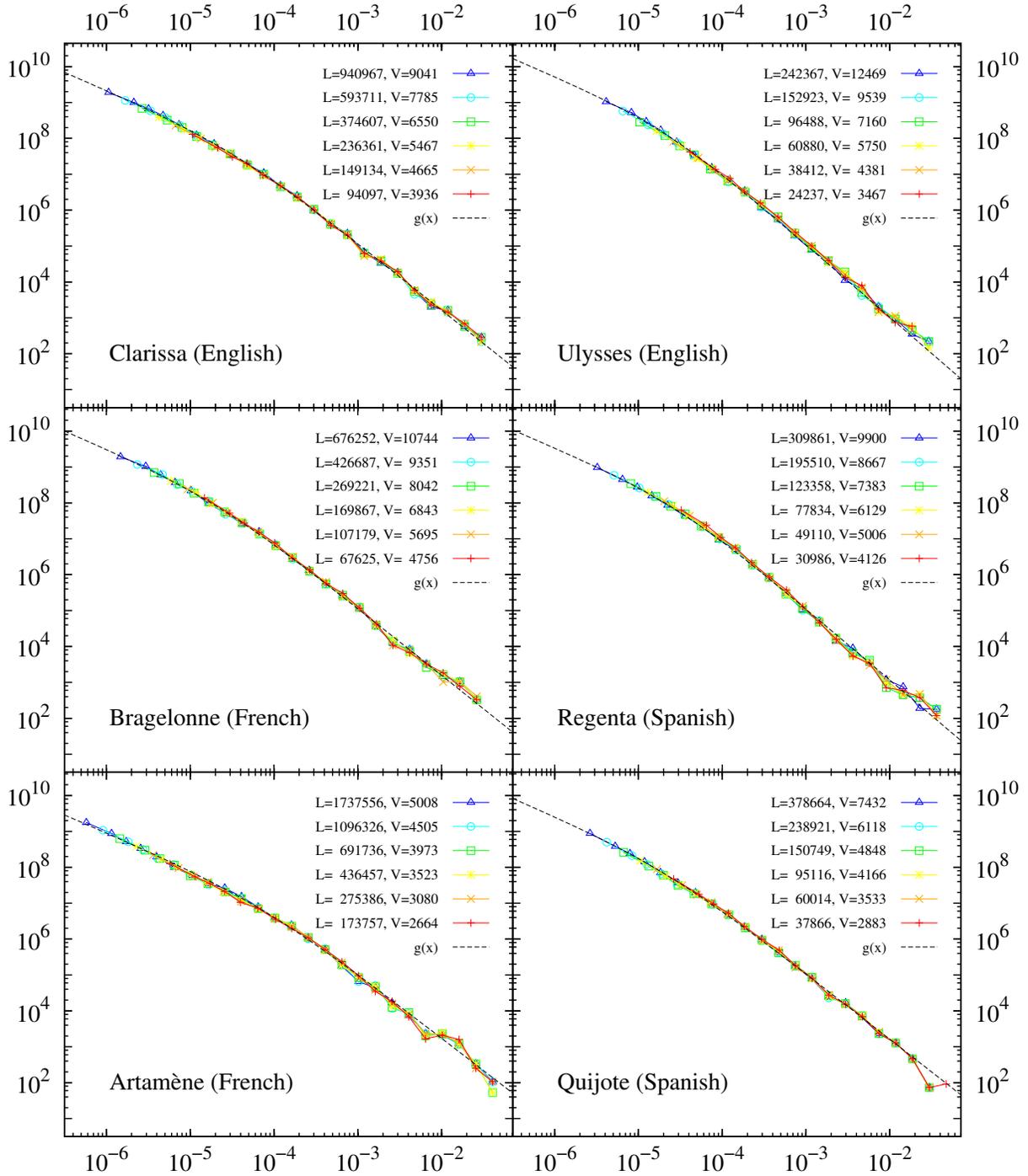}
\caption{Same rescaled distributions as in previous figure
 ($L V_L D_L(n)$ versus $n/L$), but for the frequencies of lemmas.
 The data collapse guarantees the fulfillment of the scaling law also in this case.
The fit resulting from the double power-law distribution, Eq.~\eqref{g prop}, is also included.
}
\label{data collapse lemmas}
\end{figure}

Things  turn out to be somewhat different after the lemmatization process. 
The scaling ansatz is still clearly valid for the frequency distributions, 
see Fig.~\ref{data collapse lemmas},
but with a different kind of 
scaling function $g(x)$, with 
a more defined characteristic shape, due to a more
 pronounced log-log curvature or convexity. 
In fact,
 close examination of the data leads us to conclude
that the lemmatization process enhances  the goodness
of the scaling approximation, 
specially in the low-frequency zone. 
It could be reasoned that, as lemmatized texts have a significantly reduced vocabulary 
compared to the original ones, but the total length remains essentially the same, 
they are somehow equivalent to much longer texts, if one considers the length-to-vocabulary ratio. 
Although this matter needs to be further investigated, 
it supports the idea that our main hypothesis, the scale-invariance of 
the distribution of frequencies,
holds more strongly for longer texts. 

Due to the clear curvature of $g(x)$ in the lemmatized case, 
we go one step further and propose a concrete function to fit these data, namely,
\begin{equation}
\label{g prop}
g(x) = \frac{k}{x(a+x^{\gamma-1})}.
\end{equation}
This function has two free parameters, $a$ and $\gamma$
 (with $\gamma >1$ and $a>0$), 
and behaves as a double power law,  that is, 
for large $x$, $g(x) \sim x^{-\gamma}$
(we still have Zipf's law),
while
for small $x$, $g(x)\sim x^{-1}$. 
The transition point between both power-law tails is determined by $a$
(more precisely, by $a^{\frac 1 {\gamma-1}}$), 
and $k$ is fixed by normalization. But an important issue is that it is not $g(x)$ which is normalized to one
but $D_L(n)$. 
We select a power-law with exponent one for small $x$ 
for three reasons: first, in order to 
explore an alternative to the
power law in the $V_L$ versus $L$ relation 
(which is not clearly supported by the data, see next section);
 second, to allow for a better comparison of our results and those of
Ref. \cite{Bernhardsson_2009};
third, to keep the number of parameters minimum.
 Thus, we do not look for the most accurate fit but for the
simplest description of the data.



Then, defining $n_a=a^{\frac{1}{\gamma-1}} L$, the corresponding word-frequency density 
(or, more properly, lemma-frequency density, or type-frequency density) turns out to be
\begin{equation}
\label{wfd}
D_L(n) \propto  
\frac 1
{n \left( 1+(n/n_a)^{\gamma-1} \right) },
\end{equation}
with $n_a$ the scale parameter
(recall that the scale parameter of $g(x)$ was $a^{\frac{1}{\gamma-1}}$).
The data collapse in Fig.~\ref{data collapse lemmas} and the good fit imply that
the Zipf-like exponent $\gamma$ does not depend on $L$, 
but the transition point between both power laws, $n_a$, obviously does.
Hence, 
as $L$ grows the transition to the $\sim n^{-\gamma}$ regime occurs at higher 
absolute frequencies, given by $n_a$,
but fixed relative frequencies, {given by $a^{\frac{1}{\gamma-1}}$}.
In Table~\ref{params} we report the fitted parameters for all seven books, obtained by 
maximum likelihood estimation of the frequencies of the whole books, as well as Monte Carlo estimates of their uncertainties.
 We have confirmed the stability of $\gamma$ fitting only a power-law tail
from a fixed common relative frequency, for different values of $L$ \cite{Corral_Boleda}.

\begin{table}[h]
\centering
\begin{tabular}{|l|r|c|c|}
\hline
 title  & $ n_a \pm \sigma_{n_a}$ & $ \gamma \pm \sigma_{\gamma} $&$ a \pm \sigma_a$ \\
\hline
Artam\`ene$^{(l)}$	&$ 129.7
 \pm 12.6
 $&$ 1.807 \pm 0.026
 $&$ (4.65
  \pm 0.91)\cdot10^{-4} $\\ 
\hline
 Clarissa$^{(l)}$	&$ 32.70
  \pm 2.17
  $&$ 1.864 \pm 0.021
 $&$ (1.40
  \pm 0.24)\cdot10^{-4} $\\
\hline
Don Quijote$^{(l)}$ &$ 7.91
 \pm 0.75
 $&$ 1.827 \pm 0.020
$&$ (1.35
 \pm 0.22)\cdot10^{-4} $\\
\hline
La Regenta$^{(l)}$ &$ 9.45
  \pm 0.66
  $&$ 1.983 \pm 0.021
 $&$ (3.68
  \pm 0.62)\cdot10^{-5} $\\ 
\hline
Bragelonne$^{(l)}$ &$ 14.56
 \pm 1.23
 $&$ 1.866 \pm 0.018
$&$ (9.10
 \pm 1.37)\cdot10^{-5}$\\
\hline
Moby-Dick$^{(l)}$ &$ 8.21
 \pm 0.53
 $&$ 2.050 \pm 0.024
$&$ (2.42
 \pm 0.47)\cdot10^{-5} $\\
\hline
Ulysses$^{(l)}$ &$ 
5.38 \pm 0.31
 $&$ 2.020 \pm 0.017  
$&$ ( 1.79
 \pm 0.28)\cdot10^{-5} $\\\hline 
\end{tabular}
\caption{
Values of the parameters $n_a, \gamma$, and $a$ for the lemmatized versions (indicated with 
the superscript $l$) 
of the seven complete books. 
The fits are performed numerically through maximum likelihood estimation, 
while the standard deviations come from Monte Carlo simulations,
see Appendix~\ref{appendix MLE}.}
\label{params}
\end{table}

Regarding the low-frequency exponent, one could find a better fit 
if the exponent was not fixed to be one;
however, our data does not allow this value to be well constrained.
A more important point is the influence of lemmatization errors
in the characteristics of the low-frequency regime.
Although the tools we use are rather accurate, rare words are likely to be 
assigned a wrong lemma. This limitation is intrinsic to current
computational tools and has to be considered as a part
of the lemmatization process. Nevertheless, the fact that the behavior 
at low frequencies is robust in front of a large variation in the percentage of
lemmatization errors implies that our result is a genuine consequence of the lemmatization. 
See Appendix~\ref{appendix lemmatization} for more details.

Although double power laws have been previously fit to rank-frequency plots
 for unlemmatized multi-author corpora \cite{Ferrer2000a,Altmann2012,Petersen2012}, 
the resulting exponents for large ranks (low frequencies) are different than
the ones obtained for our lemmatized single-author texts. 
Note that Ref. \cite{Altmann2012} also proposed that the crossover 
between both power laws happened for a constant number of types, 
around 7900, independently of corpus size. This corresponds indeed 
to $r=7900$ and therefore, from Eq. (\ref{rGnL}), to a fixed relative frequency.
This is certainly in agreement with our results, supporting 
the hypothesis that rank-frequency plots and
frequency distributions are stable in terms of relative frequency.

\section{An asymptotic approximation of Heaps' law}

Coming back to our scaling ansatz, Eq.~\eqref{main}, 
the normalization of $D_L(n)$ will allow us to establish a relationship
between the word-frequency distribution and the growth of the vocabulary
with text length. In the continuous approximation,
\[
1=\int_{1}^{\infty} D_L(n) dn = 
\frac{1}{V_{L}}\int_{1}^{\infty} g(n/L) \frac{dn}{L} = 
\frac{1}{V_L}\int_{1/L}^{\infty} g(x) dx = \frac{1}{V_L} G\left(\frac{1}{L}\right), 
\]
 where we have used the previous relation
 $g(x)=-G\hspace*{0.7mm}'(x)$,
and have additionally imposed $G(\infty)\equiv 0$, 
for which it is necessary that
$g(x)$ decays faster than a power law
with exponent one.
So, 
\begin{equation}
\label{Heaps}
V_L=G\left(\frac{1}{L}\right).
\end{equation}
This just means, compared to Eq. (\ref{rGnL}), that the
number of types with relative frequency greater or equal than $1/L$ is 
the vocabulary size $V_L$,
as this is the largest rank for a text of length $L$.
 It is important to notice the difference between saying that 
$G_L(1/L)=V_{L}$, which is a trivial statement, and stating that 
$G(1/L)=V_{L}$,
which provides a link between Zipf's and Heaps' law,
or, more generally, between the distribution of frequencies and the 
vocabulary growth, by approximating the latter by the former.
The quality of such an approximation will depend, of course, 
on the goodness of the scale-invariance approximation. 
In the usual case of a power-law distribution of frequencies
extending to the lowest values, 
$g(x) \propto 1/x^{\gamma}$, with $\gamma > 1$,
then $G(x) \propto 1/x^{\gamma -1}$,
which turns into Heaps' law, $V_L \propto L^{\alpha}$, with $\alpha=\gamma-1$,
in agreement with previous research
\cite{Mandelbrot61,Kornai2002,Lu_2010,Serrano_2009,Bernhardsson_2009}.



\begin{figure}[t]
\includegraphics[width=\textwidth]{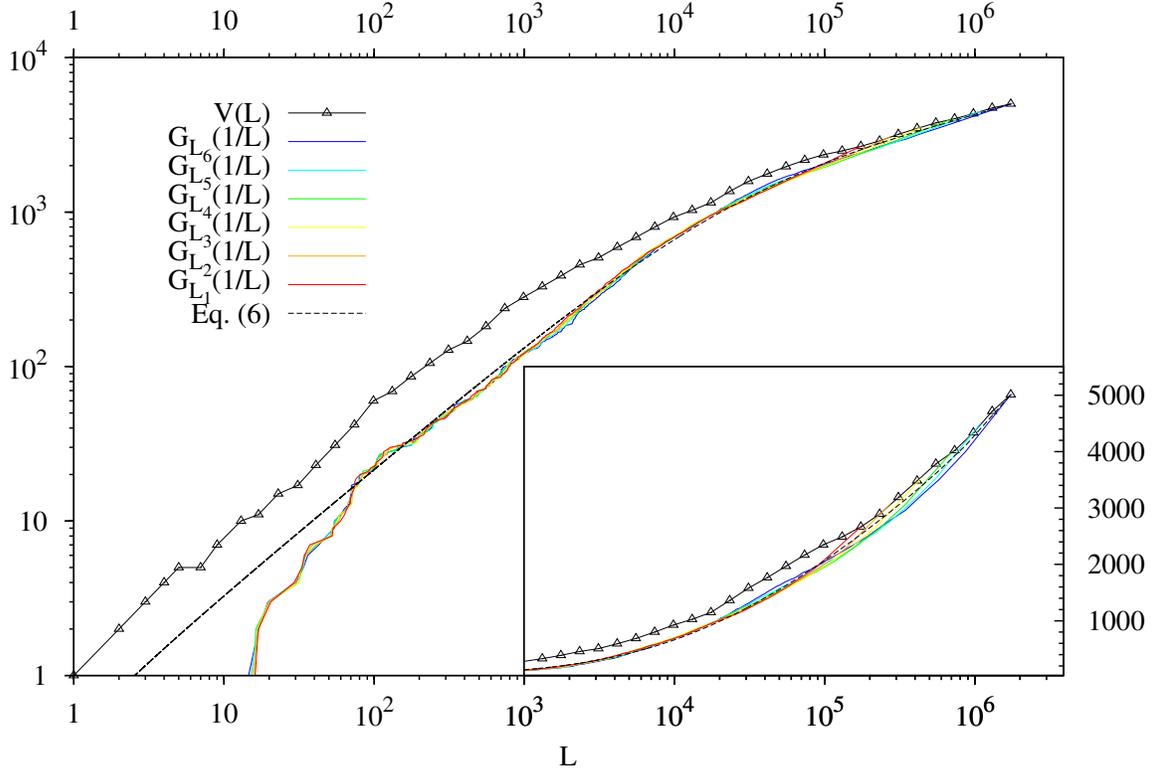}
\caption{
The actual curve $V_L$ (solid black with triangles) 
for the lemmatized version of the book \emph{Artam\`ene}, 
 together with the curves $V_L=G(1/L)$ obtained by using
the empirical inverse of the rank-frequency plot,
$r=G(n/L)$, with $L_i=L_{\text{tot}}/10^{(6-i)/5}$
(colors), 
 and the analytical expression Eq.~\eqref{discrete} with parameters determined from the fit
of $D_{L_{\text{tot}}}(n)$,  Eq.~\eqref{int g} (dashed black).
}
\label{heaps}
\end{figure}

However, this power-law growth of $V_L$ with $L$ is not what is observed in texts,
in general. 
Due to the accurate fit that we can achieve for lemmatized texts,
we can explicitly derive an asymptotic expression for $V_L$ given our proposal for $g(x)$. 
As we have just shown, $g(x)$ is not normalized to one, 
rather, $\int_{1/L}^{\infty}g(x) dx = V_L$. 
Hence, substituting $g(x)$ from Eq. (2) and integrating,
%
\begin{align}
\notag V_L =& \int_{1/L}^{\infty} \frac{k}{x(a+x^{\gamma-1})}dx
=\frac{k}{a} \int_{1/L}^{\infty}\frac{ax^{-\gamma}}{ax^{1-\gamma}+1}dx =\\
=& \frac{k}{a(1-\gamma)} \ln(a x^{1-\gamma}+1)\Bigr|_{1/L}^{\infty}
= \frac{k}{a (\gamma-1)}\ln (a L^{\gamma-1}+1).
\label{int g}
\end{align}
In this case $V_L$ is not a power law, and behaves asymptotically as $\propto \ln L$. 
This is a direct consequence of our choice for the exponent 1 in the left-tail of $g(x)$. 
Indeed, it seems clear that the vocabulary growth curve greatly deviates from a straight line 
in log-log space, for it displays a prominent convexity,
see Fig.~\ref{heaps} as an example.
Nevertheless, the result from Eq.~(\ref{int g}) is not a good fit either, due to a wrong proportionality constant. 
This is caused by the continuous approximation in Eq.~(\ref{int g}). 

For an accurate calculation of $V_L$ we must treat our variables as discrete and compute discrete sums rather than integrals. In the exact, discrete treatment of $D_L(n)$, equation~\eqref{int g} must be rewritten as
\begin{align}
\notag
V_L &= G\left( \frac{1}{L} \right)=G\left( \frac{L_{\text{tot}}/L}{L_{\text{tot}}}\right) 
= 
\sum_{n \geq L_{\text{tot}}/L} \frac{g(n/L_{\text{tot}})}{L_{\text{tot}}}=\\
&=\frac{1}{L_{\text{tot}}}\sum_{n \geq L_{\text{tot}}/L} 
\frac{k}{
\left( \frac{n}{L_{\text{tot}}} \right)
\left( a + \left( \frac{n}{L_{\text{tot}}} \right)^{\gamma-1} \right)
\label{discrete}
}
\end{align}
where we have used the fact that $S_{L_{\text{tot}}}(n')=\sum_{n \geq n'}D_{L_{\text{tot}}}(n)$,
with $n'=L_{\text{tot}}/L$
(notice that in the discrete case, $g(x)\ne -G'(x)$).
This is consistent with the fact that, indeed, the maximum likelihood parameters 
$\gamma$ and  $a$ have been computed assuming a discrete probability function 
(see Appendix~\ref{appendix MLE}), and so has the normalization constant. We would like to stress that no fit is performed in Figure~\ref{heaps}, that is, the constant $k$ in $g(x)$ is directly derived from the normalizing constant of  $D_L(n)$, and depends only on $\gamma$ and $a$.

\section{Conclusions}

In summary, we have shown
that, contrary to claims in previous research
\cite{Baayen,Powers_1998,Bernhardsson_2009},
Zipf's law in linguistics is extraordinarily
stable under changes in the size of the analyzed text.
A scaling function $g(x)$ provides a constant shape
for the distribution of frequencies of each text, $D_L(n)$,
no matter its length $L$, which only enters into the distribution
as a scale parameter and determines the size of the vocabulary $V_L$.
The apparent size-dependent exponent found previously
seems to be an artifact of the slight convexity of
 $g(x)$ in a log-log plot,
which is more clearly observed for very small values of $x$, 
accessible only for the largest text lengths.
Moreover, we find that in the case of lemmatized texts the distribution
can be well described by a double power law,
with a large-frequency exponent $\gamma$
that does not depend on $L$,
and a transition point $n_a$
that scales linearly with $L$.
The small-frequency exponent
is different
than the ones reported in Refs. \cite{Ferrer2000a,Altmann2012} 
for non-lemmatized corpora.
 Further, the stability of the shape of the frequency distribution 
allows one to predict the growth of vocabulary size with text length, 
resulting in a generalization of the popular Heaps' law.

The robustness of Zipf-like parameters under changes in system size
opens the way to more practical applications of word statistics.
In particular, we provide a consistent way to compare statistical 
properties of texts with different lengths \cite{Ferrer2013a}.
Another interesting issue would be the application of the same scaling methods
to other fields in which Zipf's law has been proposed to hold,
as economics and demography, for instance.\\




%

\appendix\section{Lemmatization}
\label{appendix lemmatization}

To analyze the distribution of frequencies of lemmas,
the texts needed to be lemmatized. 
To manually lemmatize the words would have exceeded the possibilities of this project, 
so we proceeded to automatic processing with standard computational tools: 
{\it FreeLing} \cite{Freeling} for Spanish and English
and
{\it TreeTagger} \cite{TreeTagger} for French.
The tools carry out the following steps:

\begin{enumerate}

\item 	Tokenization: 
Segmentation of the texts into sentences and sentences into words (tokens).

\item
Morphological analysis: Assignment of one or more lemmas and morphological information
(tag) to each token. For instance, \textit{found} in English can correspond to the past tense of the verb \textit{find} or to the base form of the verb \textit{found}. At this
stage, both are assigned whenever the word form \textit{found} is encountered.

\item
Morphological disambiguation: An automatic tagger assigns the single most probable lemma
and tag to each word form, depending on the context. For instance, in \textit{I found the keys} the tagger would assign the lemma \textit{find} to the word \textit{found}, while in \textit{He promised to found a hospital}, the lemma \textit{found} would be preferred.
\end{enumerate}


All these steps are automatic, such that errors are introduced at each step. 
However, the accuracy of the tools is quite high 
(e.g., around 95-97\% at the token level for morphological disambiguation), 
so a quantitative analysis based on the results of the automatic process can be carried out.
Also note that step 2 is based on a pre-existing dictionary 
(of words, not of lemmas, also called a lexicon): 
only the words that are in the dictionary 
are assigned a reliable set of morphological tags and lemmas. 
Although most of the tools used heuristically assign tag and/or lemma 
information to words that are not in the dictionary, 
we only count tokens of lemmas for which the corresponding word types are found in the dictionary, 
so as to minimize the amount of error 
introduced by the automatic processing. 
This comes at the expense of losing some data. 
However, the dictionaries have quite a good coverage of the vocabulary, particularly at the token level, 
but also at the type level (see Table \ref{Tablethree}). 
The exceptions are {\it Ulysses}, because of the stream of consciousness prose, 
which uses many non-standard word forms, and {\it Artam\`ene}, 
because 17th century French contains many word forms 
that a dictionary of modern French does not include.

\begin{table}[!ht]
\caption{
Coverage of the vocabulary by the dictionary in each language, 
both at the type and at the token level.
{Remember that we
distinguish between a word {\it type} (corresponding to its orthographic form) 
and its {\it tokens} (actual occurrences in text). 
}
}
\begin{tabular}{|lrr|}
\hline
 title         & types     & tokens   \\
\hline
Clarissa       & 68.0 \%  & 96.9 \% \\
Moby-Dick      & 70.8 \%  & 94.7 \% \\ 
Ulysses        & 58.6 \%  & 90.4 \%  \\
Don Quijote    & 81.3 \%  & 97.0 \%  \\
La Regenta     & 89.5 \%  & 97.9 \%  \\
Artam\`ene     & 43.6 \%  & 83.6 \%  \\
Bragelonne     & 89.8 \%  & 97.5 \%  \\
Seitsem\"an v. & 89.8 \%  & 95.4 \%  \\
Kev\"at ja t.  & 96.2 \%  & 98.3 \%  \\
Vanhempieni r. & 96.5 \%  & 98.5 \%  \\
\hline
average        & 78.4 \%  & 95.0 \%  \\
\hline
\end{tabular}
\label{Tablethree}
\end{table}

\section{Maximum likelihood fitting}
\label{appendix MLE}

The fitted values of Table~\ref{params} have been obtained by maximum-likelihood estimation  (MLE). 
This well-known procedure consists firstly in computing the log-likelihood function $\mathcal{L}$, which in our case reads

\[
\mathcal{L}=\frac{1}{V_L}\sum_{i=1}^{V_L}\ln D_{L}(n_i) 
=\ln K
- \frac{1}{V_L}\sum_{i=1}^{V_L} \ln \left(n_i(b+n_i^{\gamma-1})\right)
\]
with 
 $n_i$ the $V_L$ values of the frequency
and
the normalization constant $K$ in the discrete case equal to

\[
K=\left[\sum_{n=1}^{n_{\max}} \frac{1}{n(b + n^{\gamma -1})}\right]^{-1}.
\]
Note that we have reparameterized the distribution compared to the main text, 
introducing $b=n_a^{\gamma-1}=a L^{\gamma-1}$.
Then, $\mathcal{L}$ is maximized with respect to the parameters $\gamma$ and $b$; 
this has been done numerically using the simplex method \cite{Press}. 
The error terms $\sigma_{\gamma}$ and $\sigma_{b}$,
representing the standard deviation of each estimator, 
are computed from Monte Carlo simulations: 
From the resulting maximum-likelihood parameters $\gamma^{*}$ and $b^{*}$, 
synthetic data samples are simulated, and the MLE parameters of these samples
are calculated in the same way; their fluctuations yield $\sigma_{\gamma}$ and $\sigma_{b}$.
 We stress that no continuous approximation has been made, that is, the simulated data follows the discrete probability function $D_L(n)$ 
(this is done using the rejection method, see Ref. \cite{Devroye,Corral_Boleda} for details for a similar case). In a summarized recipe, the procedure simply is:
\begin{enumerate}
\item Numerically compute the MLE parameters, $\gamma^*$ and $b^*$.
\item Draw $M$ datasets, each of size $V_L$, from the discrete probability function $D_{L}(n; \gamma^*,b^*)$.
\item For each dataset $m=1\dots M$, compute the MLE parameters $\gamma^{m},b^{m}$.
\item Compute the standard deviations $\sigma_\gamma$ and $\sigma_{b}$ of the sets $\{\gamma^m\}_{m=1}^M$ and $\{b^m\}_{m=1}^M$.
\end{enumerate}
The standard deviations of $n_a$ and $a$ are computed in the same way using
their relationship to $b$ and $\gamma$.

\acknowledgments

We appreciate a collaboration with R. Ferrer-i-Cancho,
who also put A. C. in contact with G. B.
Financial support is acknowledged from grants FIS2009-09508 from the Ministerio de Ciencia y Tecnolog\'{\i}a,
FIS2012-31324 from the Ministerio de Econom\'{\i}a y Competitividad,
and 2009-SGR-164 from Generalitat de Catalunya, which also supported 
F. F.-C. through grant 2012FI\_B 00422 and
G. B. through AGAUR grant 2010BP-A00070.

\bibliographystyle{apsrev}
\bibliography{biblio}

\end{document}